\documentclass[12pt]{article}

\usepackage{lineno}
\modulolinenumbers[5]
\usepackage{caption}
\usepackage[a4paper]{geometry}
\usepackage{amsfonts}
\usepackage{arydshln}
\usepackage{xtab}
\usepackage{setspace}
\usepackage{fancyhdr}
\usepackage[dvips]{graphicx}
\usepackage[T1]{fontenc}
\usepackage[latin1,utf8]{inputenc}
\usepackage[main=english,italian]{babel}
\usepackage{type1ec}
\usepackage[final]{microtype}
\usepackage{amsmath,amssymb,amsthm,eucal,dsfont,bm,mathrsfs,stmaryrd}
\usepackage{lmodern}
\usepackage{textcomp}
\usepackage{pict2e,enumitem}
\usepackage{hyperref}
\usepackage[dvipsnames]{xcolor}
\usepackage[nodayofweek]{datetime} 
\usepackage{comment}

\hypersetup{
    pdfpagemode={UseOutlines},
    bookmarksopen,
    pdfstartview={FitH},
    colorlinks,
    linkcolor={purple},
    citecolor={cornellred},
    urlcolor={black}
            }

\definecolor{cornellred}{rgb}{0.7, 0.11, 0.11}

\usepackage{geometry}

\newdateformat{monthyear}{\monthname[\THEMONTH], \THEYEAR}

\theoremstyle{plain}

\theoremstyle{definition}
        
        \newtheorem*{remark}{Remark}
        \newtheorem{example}{Example}[section]
        
\renewcommand
        {\thefootnote}{\arabic{footnote}}
        
\newcommand{\symfootnote}[1]{%
\let\oldthefootnote=\thefootnote%
\stepcounter{mpfootnote}%
\addtocounter{footnote}{-1}%
\renewcommand{\thefootnote}{\fnsymbol{mpfootnote}}%
\footnote{#1}%
\let\thefootnote=\oldthefootnote%
}

\usepackage{tikz}

\newcommand\blfootnote[1]{%
  \begingroup
  \renewcommand\thefootnote{}\footnote{#1}%
  \addtocounter{footnote}{-1}%
  \endgroup
}

\makeatletter
\def\bbibitem#1{\item[]%
    \if@filesw\immediate\write\@auxout{\string \bibcite {#1}{\the\value{\@listctr }}}\fi\ignorespaces}
\makeatletter

\title{Group Fairness Is Not Derivable From Justice: \\a Mathematical Proof}
\author{Nicolò Cangiotti\footnote{Politecnico di Milano, via Bonardi 9, Campus Leonardo, 20133, Milan (Italy). E-mail: \texttt{nicolo.cangiotti@polimi.it}} \ \& Michele Loi\footnote{Politecnico di Milano, via Bonardi 9, Campus Leonardo, 20133, Milan (Italy). E-mail: \texttt{michele.loi@polimi.it}}}

\date{}

\begin{document}

\maketitle

\begin{abstract}
We argue that an imperfect criminal law procedure cannot be group-fair, if “group fairness” involves ensuring the same chances of acquittal or convictions to all innocent defendants independently of their morally arbitrary features. We show mathematically that only a perfect procedure (involving no mistake), a non-deterministic one, or a degenerate one (everyone or no one is convicted) can guarantee “group fairness”, in the general case. Following a recent proposal, we adopt a definition of “group fairness”, requiring that individuals who are equal in merit ought to have the same statistical chances of obtaining advantages and disadvantages, in a way that is statistically independent of any of their feature that does not count as merit. We define justice as everyone obtaining advantages and disadvantages that they merit, where merit (or desert) is defined as that feature (or combination of features) of individuals, if any, that justifies the unequal distribution of advantages or disadvantages between them. The circularity in this definition is intended, as it points to a general conflict between group fairness and justice that obtains independently of substantive views about the justification of inequality (e.g., whether inequality is justified by unequal needs, responsibilities or contributions). We explain by mathematical argument that the only imperfect procedures offering an a-priori guarantee of fairness in relation to all non-merit trait are lotteries or degenerate ones (i.e., everyone or no one is convicted). To provide a more intuitive point of view, we exploit an adjustment of the well-known ROC space, in order to represent all possible procedures in our model by a schematic diagram. The argument seems to be equally valid for all human procedures, provided they are imperfect. This clearly includes algorithmic decision-making, including decisions based on statistical predictions, since in practice all statistical models are error prone.
\blfootnote{Both authors contributed equally to this research.}
\bigskip

\end{abstract}

\section{Introduction}
\label{Sec1}
The main aim of this paper is, after introducing a distinction between justice and group fairness through the language of probability, to show that theories of justice do not provide a sufficient normative grounding for reasonable accounts of group fairness. Justice and fairness may be used as rough synonyms when it is not assumed that they stand for different concepts.\footnote{For example, the same luck-egalitarian view is described as "comparative fairness" in \cite{temkin_equality_2017} and as "egalitarian justice" in \cite{cohen_currency_1989}.} Remarkably, one important theory, John Rawls's, allegedly derives justice from a combination of formal fairness and prudential premises. Our starting point is different: group fairness and justice are compatible only when justice is perfectly realized, otherwise one obtains at the expense of the other. And yet, if one tries to define \emph{fairness} with the same normative elements that are used to define justice, it turns out that group-fairness can only be achieved in an absolute sense by procedures involving a non-deterministic element. This interesting result could be regarded as an insight about the impossibility of group fairness; but it could also be regarded as a reason to define group-fairness on the basis of theories \emph{other} than theories of justice.

When we talk about fairness, here, we mean a property of procedures allocating outcomes, not Platonic properties of outcome distributions defined independently of the procedural elements that may bring those distributions about. We focus on \emph{imperfectly just} procedures, those that do not guarantee a perfectly just distribution. The property of group fairness as a property of procedures, in the definition from which we start, is an intuitive one. It seems intuitive that a procedure is group fair only if it does not favor in a morally arbitrary way, intentionally or unintentionally, any individual who belongs to a group over an individual who belongs to a different group. We will subsequently provide a rigorous analysis of this notion, following a notation defended for the first time in \cite{loi_philosophical_2019}. We then provide a formal argument to show that, for an imperfectly just procedure to be group fair, it must necessarily be a non-deterministic one.

As procedures involving a non-deterministic element are typically perceived as problematic from the point of view of justice, this result is both unexpected and not trivial. We examine the relation between outcome justice and procedural justice based on \cite{rawls_theory_1971} the distinction between pure, perfect, and imperfect procedural justice. We exclude pure procedural justice where just outcomes are not defined prior to procedures. Then, we explain the relation between (perfect and imperfect) procedural justice and group fairness in the language of probability (and set) theory. In this way, we can show mathematically that, among deterministic procedures, only a perfect procedure can be group-fair with respect to all groups. We conclude that, for all imperfect procedures, only non-deterministic ones can be group fair with respect to all (logically possible) groups. The statement we prove mathematically is: \emph{unless the procedure is perfect, one can always identify at least two morally arbitrary groups, relative to which the procedure is not fair.} This result can be shown mathematically, given our definitions of justice and group fairness. The only way to avoid this result is to make the procedure non-deterministic.

This result is relevant for the FAccT community because many algorithms qualify as procedures according to our definition. It also applies to many procedures that are not currently discussed in the context of data science. This requires a new, more abstract language, that is an essential element of this (and similar, see \cite{loi_philosophical_2019}) contributions.

Furthermore, with a stark contrast to papers discussing statistical fairness definitions for machine learning, our goal is not to deliver an measurable definition of group fairness. Our argument illustrates the relevance of postulating an objective \emph{moral} ground truth for the sake of the normative analysis. Our basic philosophical postulate is that someone may be objectively deserving even if no feasible procedure for determining who deserves what in real life exists. For example, someone could be objectively innocent of the crime charged to her, and deserve acquittal, on the basis of evidence that a procedure, even the best human procedure, may ignore, which leads to an (unjust) conviction. Perfect justice is consistent as a Platonic idea, even when we know no procedure that may achieve it. The concept of the moral ground truth differs from the concept of the ground truth as used in the machine learning literature, when it typically refers to features that are actually observed (e.g., arrest by the police) and whose normative relevance is in many cases worth doubting. Our paper needs a different conception of the ground truth because its point is, primarily, philosophical and conceptual.

The structure of the paper is the following. Section \ref{Related} discusses related work. Section \ref{Concepts} provides definitions of procedure, merit, morally arbitrary group, justice, fairness, perfect and imperfect (as attributes of procedures). Section \ref{Group fairness} offers a schematic and graphical representation of all possible procedures for distributing advantages or disadvantages in a context in which some inequalities are morally justified; we provide this representation only for the binary case for simplicity’s sake; the representation relies on probability.  Section \ref{Argument1} presents briefly the main argument of our analysis. In Section \ref{Argument2} we examine deeply the argument by considering all possible procedures. Section \ref{Discussion} is devoted to the final discussion, and Section \ref{Conclusions} takes stock of our work.

\section{Related work}
\label{Related}
The distinction between a perfect and imperfect procedure derives from John Rawls \cite{rawls_theory_1971}, who defines perfect procedural justice as a form of procedural justice guaranteeing the justice of outcomes, where outcomes are characterized as just prior to the procedure. Imperfect procedural justice results when just outcomes cannot be guaranteed every time.

In the contemporary debate in analytical philosophy (e.g. \cite{arneson_equality_1989,miller_principles_1999,sen_equality_1982}) the goal is, often, to characterize outcomes that are just in this sense. For example, luck egalitarians \cite{arneson_equality_1989} characterize just inequalities as inequalities that reflect unequal responsibility; "desertarians" as outcomes reflecting unequal contributions \cite{brouwer_why_2018}. One may also argue that different needs justify inequality \cite{herlitz_measuring_2016}. Pluralists \cite{miller_principles_1999, walzer_spheres_1983} maintain that inequalities are justified by different properties in different contexts. 

Philosophers interested in procedural justice mostly discuss \emph{pure} procedural justice, where a prior, substantive definition of just outcomes is not available, or subject to (sometimes, reasonable) disagreement \cite{ceva_interactive_2016,daniels_accountability_2000,wong_democratizing_2019}. Here what makes outcomes just \textit{is} the procedure leading to them. We ignore pure procedural justice in what follows. 

Our definition of "group fairness" bears resemblance to \cite{hardt_equality_2016}'s definition of equalized odds (and "separation" \cite{barocas_fairness_nodate}). Equalized odds requires that, when testing an algorithm with historical data, individuals with the same value for the ground truth, $Y$ (e.g., those who have repaid their loan), have the same chances of receiving a favorable or unfavorable classification, $\hat{Y}$ (e.g., they are classified as future defaulting creditors), independently of their group. This is not necessarily "fair", according to our use of the term, because equalized odds are defined relative to the actually observed ground truth, which may differ from the moral ground truth, which is what morally justifies the unequal treatment \cite{heidari_moral_2019}.

The definition of group fairness we analyze is not entirely new. The normative definition we use here has been explicitly defended \cite{loi_philosophical_2019,loi_fair_2021}, or, more often, presupposed. For example, in \cite{di_bello_profile_2020}, one premise of the argument is that equal protection implies "a right of innocent defendants not to be exposed to higher ex ante risks of mistaken conviction compared to other innocent defendants facing similar charges" (p. 147). In the vocabulary we employ here, this amounts to assuming that all (objectively) innocent defendants equally deserve to be acquitted, independently of their statistically relevant characteristics. Our interpretation is supported by the objection against statistical profiling being that "admitting incriminating profile evidence would create an inequality in the distribution of the risks of mistaken conviction [between innocent individuals matching and not matching statistical profiles], and that admitting exculpatory profile evidence would do so as well" (p. 167). 

The equality of opportunity approach, also discussed in relation to machine learning \cite{heidari_moral_2019} rests on the luck-egalitarian assumption that a person's effort (that for which the individual is responsible) justifies inequality; by contrast, his or her circumstances, (e.g, "the socio-economic status he/she is born into") do not. We use "merit" here in a way compatible with the (extensional) identity between "effort", as used in an equality of opportunity theory of this type \cite{heidari_moral_2019} and merit; but our concept is different (intensionally). It denotes different properties if other theories are valid, or in different distributive contexts.\footnote{We follow \cite{loi_philosophical_2019,loi_fair_2021}, who use "desert*" to designate any inequality-justifier, which corresponds to different properties according to responsibility-based views (e.g., \cite{heidari_moral_2019}, meritocratic, \cite{brouwer_why_2018, miller_principles_1999}, need-based or pluralistic ones \cite{miller_principles_1999}).}

\section{Main concepts}
\label{Concepts}
\subsection{Procedure and utility}
We define a procedure as a rule leading to the allocation of benefits, burdens or harms of various kinds. Benefits and harms are (or rather become, by virtue of the procedure) features of the individuals obtaining them. we indicate these with the letter $U$.  

More precisely, by a procedure we mean a sequence of actions such that, when criterion $X$ is satisfied, a given outcome for the individual, $U$, is produced. For the sake of simplicity we prove our argument for procedures leading to binary decisions, e.g., an individual is either convicted or acquitted in a trial, hired or rejected by an employer, given or refused bail, etc. Some procedures are extremely complicated. They involve a large number of criteria that jointly determine a decision. But we can suppose for the sake of simplicity that $X$ is a uber-criterion which corresponds to all a manifold of other criteria, or combination of criteria, being satisfied, to such an extent that their satisfaction is sufficient and necessary for the decision concerning the individual. The nature of $X$ may be very complicated; in some cases, $X$ combines several criteria of different weights, neither of which is individually necessary or sufficient for a decision. 

We assume, for deterministic procedures, a set of "determinant facts" $X$ that summarizes, for that procedure, the satisfaction of all the criteria that are relevant and every other fact about the circumstances affecting the procedure explaining why a given outcome will necessarily be achieved. Take the criminal trial for instance. In the criminal trial there are many sources of evidence that have to be presented to a court and evaluated, in order for the judge or the jury to reach a decision. Many witnesses have to be consulted, different types of evidence and epistemic criteria have to be balanced, etc. Moreover, most procedures are probably sensitive to inputs that they were not \emph{intended} to be responsive to. These may include the way the defendant expresses his or her emotions when speaking, the temperature of the room on the trial day, whether a judge is hungry, the color of the skin of the defendant and the witnesses, etc. We shall refer to these as "procedure-influencing circumstances". We simplify this picture by supposing the decision to be determined by a set of determining causes, the existence of which we represent by the value of the binary random variable $X$. 

Procedural changes may imply different outcomes for different type of defendants, though the effects may not be directly measurable. If a new \emph{procedure 2} is adopted, innocent defendants may have a higher chance of being acquitted of a given charge, than under old procedure 1. Unfortunately, it may also be the case that guilty defendants are also less likely to be convicted as a result of procedure 2. Moreover, different procedures typically imply that a different set of determinant facts $X=1$ is necessary and sufficient to acquit a defendant. In other words, what facts count as necessary and sufficient conditions for a acquittal under procedure 1 may not count as such under procedure 2. Moreover, members of one demographic group (e.g., young people) may have higher chances of being acquitted under procedure 3 compared to procedure 2, even though innocent defendants are equally likely to be acquitted in both.

We can also describe the two possible outcomes for individuals subjected to a trial as unequal utility outcomes for them, e.g., a better outcome, $U=1$, if the individual is acquitted, and a worse outcome, $U=0$, if he is convicted.

Thus, we can summarize all the salient events taking place in a criminal trial procedure by the following definitions:
\begin{itemize}
    \item $X=1$ the determinant facts for acquittal are in place;\footnote{One may think about $1$ as lack of (perceived) sufficient evidence for the crime.}
    \item $X=0$ the determinant facts for acquittal are not in place;
    \item $U=1$ the defendant is released of all charges and acquitted;
    \item $U=0$ the defendant is declared guilty and convicted.
\end{itemize}

The necessary and sufficient condition of our criterion with respect the benefit can be summarized by the following expression:

\begin{equation}
\label{XU}
X=1 \iff U=1,
\end{equation}

which (in the binary case) amounts to

\begin{equation*}
X=1 \Rightarrow U=1 \wedge X=0 \Rightarrow U=0.
\end{equation*}
In the following we shall introduce a probabilistic notation, which should help to formalize mathematically our argument. In particular we use the classical notation $\mathbb{P}(\cdot)$ for the probability and $\mathbb{P}(\cdot \ \vert \ \cdot )$ for the conditional probability. In these terms, a procedure can be see as a the method to associate the outcome $U$ with a couple of probabilities: 
\begin{align*}
\mathbb{P}(U=0)&=p_{0};\\
\mathbb{P}(U=1)&=p_{1}.
\end{align*}
It is trivial to observe that in the binary case we have $p_1=1-p_0$ and the second equality is directly deductible from the first one. This definitions build a close link between the procedure and its probabilistic description. By using relation \eqref{XU}, we can also find the probability relations for the (binary) criterion $X$:
\begin{align*}
\mathbb{P}(X=0)&=p_{0};\\
\mathbb{P}(X=1)&=p_{1}.
\end{align*}
Moreover, we notice that the following conditional probabilities hold:
\begin{align*}
\mathbb{P}(U=0\ |\ X=0)=1;\\
\mathbb{P}(U=0\ |\ X=1)=0;\\
\mathbb{P}(U=1\ |\ X=1)=1;\\
\mathbb{P}(U=1\ |\ X=0)=0.
\end{align*}

Let us reflect now on the ontology of the determinant facts $X$. If a procedure is deterministic, there will be a set of inputs (in advance of the procedure being actually carried out) that always lead to the same output, e.g., a conviction. At least in the case of purely mechanical (unambiguously-defined) deterministic procedures with two mutually exclusive outcomes, (for example, acquittal and conviction), $X$ can be regarded as the \emph{universal unambiguous criterion} (in what follows, criterion) that necessarily leads to the procedure output.\footnote{\label{note}In logical words, we are considering a certain true/false decision problem, which is \emph{decidable}. A good example could be the formalization of a given procedure with a Turing-decidable algorithm.} (This is intended to make our argument especially clear, even if it is, morally speaking, an unnecessarily restrictive assumption.) For example, consider a criminal law procedure which takes, as inputs, a description of the case by the defendant's lawyer, another one by the prosecutor, counts the sum of characters in both texts,  $n$, and convicts the defendants if and only if $n$ is prime. It is intuitively clear that the procedure reaches a determinate verdict with a specific result for every combination of inputs and that this result will be reached in a purely mechanical manner. In this case, we can ascribe to the random variable $X$ the value $1$ if $n$ \emph{is prime}, $0$ otherwise. We can then consider the group of all individuals for which $X=1$. We shall say that these are the individuals for which the criterion for acquittal is satisfied. Notice that, for very large reports, it may not be immediately clear whether $X=1$ before the procedure (the computations needed to determine if \emph{n} is prime) is actually carried out. Yet, the determinant facts exist independently of carrying out the procedure since the very moment the two reports are completed.

The concept of determinant criterion can be generalized to fully deterministic procedures that are not unambiguous like the mechanical procedure imagined here. In the real world, the "official" inputs, those that are intended to determine a trial (e.g., the lawyers' arguments and witnesses' depositions), do not cause the outcomes with full certainty, but only affect their probabilities. The full set of causes producing an acquittal outcome include the (typically, unintended) procedure-affecting circumstances, such as the defendants' tone of voice or the witness skin color, mentioned before. Yet there may be a level of description at which even real world procedures are deterministic. The meaning of "deterministic" for procedures can be described in modal terms: given a sufficiently exhaustive description of the initial state of the procedure (including the procedure-affecting circumstances), necessarily the same outcome is produced, at least in suitably defined normal operating conditions. For example, if a judge must interpret the laws rather than apply them mechanically, the same judge would not change the interpretation given the same (suitably described) initial state of the procedure.\footnote{To change interpretation without any change in the circumstances of the case would amount to introduce a degree of randomness in its normal operations.} The set of determinant causes $X$ is, unlike the unambiguous criterion of a mechanical procedure, typically too complex for any human to describe.\footnote{A careful reader may notice that the set of inputs should be fixed. However in the real world laws are subject to different interpretations, which could make the set of inputs not definable in advance. The formal proof relies on a mathematical construction that needs some strong hypothesis on $X$ as well as a procedure, which acts as an algorithm, see also Note \ref{note}.} 

\subsection{Justice, deserved and undeserved inequalities}
Following \cite{loi_philosophical_2019}, we assume the existence of a theory of justice for the allocation of (dis)advantages, denoted with values of the random variable $U$, in the population.  A theory of outcome justice is a theory of what, if anything, justifies inequalities in the distribution of $U$. We refer to such "inequality justifier" as "desert" or "merit" without any intended implication that we are endorsing a substantive meritocratic view. We abstain from taking any position on the substantive philosophical problem of what, if anything, justifies inequality.  Notice that this definition accounts for justice in terms of the distribution of a type of outcome (denoted with $U$), given other features of individuals (e.g., their contributions or needs, i.e., $J$); this completely ignores the procedures that have been used to allocate $U$ in practice.

We use $J$ for a random variable, the values of which indicate features, among those of an individual, making that individual deserving of a certain treatment, e.g., a person's need for welfare assistance, her excellence for a prize, her responsibility of a crime for the punishment of that crime, etc.

We assume that, given a context (e.g., the distribution of convictions, or health-care resources), such values of $J$ can be defined. One could argue that only one feature (e.g., responsibility) justifies inequality in all contexts, but our proof does not need this assumption.  Our analysis is conceptually independent from a specific justifier, as long as it can be abstractly described in this abstract vocabulary. We suppose that even complex accounts of merit can be described as distinct values of a single variable $J$ (which may be thought as a vector of several features). We simply assume that, no matter how complex the inequality-justifying property, individuals will differ in relation to it in an objective way. 

For the sake of simplifying our proof, we consider here a case in which $J$ is binary, e.g., there are only two classes of people who differ in relation to merit, that is to say, people either deserve a positive outcome ($J=1$) or not ($J=0$). The argument generalizes unproblematically to a different range of values for $J$.
 
In our example, we assume that defendants who are guilty of a crime deserve their conviction and defendants who are not guilty deserve clearing of all charges.

We define inequalities in the distribution of $U$ among individuals who are equal in $J$ as undeserved inequalities, or equivalently, unjust (but not necessarily unfair) inequalities. For example, consider two people both guilty of the crime charged to them and suppose that one is convicted and the other not. The inequality between these two people is deserved, because the two people differ in their desert, $J$. Conversely, the inequality between two defendants who are both innocent (or guilty) is not deserved. If a procedure were perfect, as we shall see next, all equally deserving individuals would receive the same outcomes. For imperfect procedures, it may be probable that equally deserving individuals are allocated equal (dis)advantages, but it is never certain. This uncertainty can be expressed as a probability. Thus, for this class of individuals $J=1$, and $X=0$.

We shall introduce the following notations for the development of our mathematical argument:
\begin{itemize}
    \item $J=1$ the person is (really) innocent;
    \item $J=0$ the person is (really) guilty.
\end{itemize}

We can introduce now the conditional effect of the $J$ in the description of the procedure by considering the following four probability:
\begin{align*}
\mathbb{P}(U=0  \ | \ J=1)&=p_{01};\\
\mathbb{P}(U= 0  \ | \ J=0)&=p_{00};\\
\mathbb{P}(U= 1 \ | \ J=1)&=p_{11};\\
\mathbb{P}(U= 1 \ | \ J=0)&=p_{10}.
\end{align*}
As explained above, the binary case allow us to limit our analysis on the first two equations without loss of information.

Notice that we distinguish between $J$, which makes an individual deserving of $U$, and $X$, which determines that the individual will be assigned $U$ by the procedure. In the criminal justice case, for example, $X=0$ includes the evidence submitted to the court, while $J=0$ is the fact that the defendant is actually guilty, including facts for which no court evidence can be produced. When a procedure is imperfect $J$ and $X$ do not correspond. For example, in the case of some innocent defendants ($J=1$), her features and circumstances are sufficient to cause the appearance of guilt ($X=0$), leading to a conviction ($U=0$). A case such as this would be formalized as $J=1$, and $X=0$, $p_{01}=1$ and $p_{10}=0$.

\subsection{Morally arbitrary groups}
Given a background theory of justice which we use to determine the nature of $J$, we define morally arbitrary traits as those features the individual has which do not justify inequalities. Thus, every feature that is not a possible value of $J$ is a morally arbitrary feature. Individuals who are equal in $J$ (deserve the same $U$) and differ in other respects differ in a morally arbitrary way, relative to the good $U$, whose distribution is in question. We appeal to the concept of morally arbitrary features in order to assess if a procedure, described probabilistically, exhibits any degree of group-unfairness. We illustrate this with the following example.

\begin{example}
\label{exampleMF}
Let us consider a population of $12000$ individuals. Moreover, let us consider the classical binary attributes male/female ($\{M,F\}$). We notice that one can also describe this kind of situation in terms of "sex" $S$, with $S\in\{0,1\}$, for simplicity, in the following lines we shall use $\{M,F\}$. We assume that a person's sex is not, in itself, a good ground to acquit or convict individuals. Hence, sex is a morally arbitrary feature.
Let now suppose an hypothetical division of the population as follows.
\smallskip

\begin{center}
\begin{tabular}{|c|c|}
\hline
\textbf{Attribute}& \textbf{Individuals}\\
\hline
Male ($M$)& $6000$ \\
\hline
Female ($F$) & $4000$ \\
\hline
\end{tabular}
\end{center}
\smallskip

Let us suppose that a \emph{dataset} provides the following \emph{table of guilt} and its complement.
\smallskip

\begin{center}
\begin{tabular}{|c|c|}
\hline
\multicolumn{2}{|c|}{\textbf{Guilty} (${J=0}$)}\\
\hline
\textbf{Attribute}& \textbf{Individuals}\\
\hline
Male ($M$)& $2000$ \\
\hline
Female ($F$) & $500$ \\
\hline
\end{tabular}
\qquad
\begin{tabular}{|c|c|}
\hline
\multicolumn{2}{|c|}{\textbf{ Not Guilty} (${J=1}$)}\\
\hline
\textbf{Attribute}& \textbf{Individuals}\\
\hline
Male ($M$)& $4000$ \\
\hline
Female ($F$) & $3500$ \\
\hline
\end{tabular}
\end{center}
\smallskip

Now let we start from a given procedure such that
\begin{align*}
\mathbb{P}(U=0 \ | \ J=0)&=3/4;\\
\mathbb{P}(U=0 \ | \ J=1)&=1/10. 
\end{align*}
In this toy model, the number of convicted (i.e., $U=0$) is
\smallskip

\begin{center}
\begin{tabular}{|c|c|}
\hline
GUILTY CONVICTED & $1875$ \\
\hline
NOT GUILTY CONVICTED & $750$ \\
\hline
\end{tabular}
\end{center}
\smallskip

Now, let us suppose that, as a requirement of group-fairness, we should guarantee that innocent individuals have the same chances of acquittal and guilty individuals the same chances of conviction, in either case, independently of their sex. (See \ref{GroupFair} for the formal definition of group-fairness.) Let us suppose that, after optimizing the procedure for accuracy constrained by group fairness, we obtain the following results:
\begin{align*}
\mathbb{P}(U=0 \ | \ J=1 \wedge M) =& \ \mathbb{P}(U=0 \ | \ J=1 \wedge F)=  3/4;\\
\mathbb{P}(U=0 \ | \ J=0 \wedge M)=& \ \mathbb{P}(U=0 \ | \ J=0 \wedge F)= 1/10,
\end{align*}
leading to the following two tables:

\begin{center}
\resizebox{15cm}{!}{
\begin{tabular}{|c|c|}
\hline
MALE GUILTY CONVICTED & $1500$ \\
\hline
MALE NOT GUILTY CONVICTED & $400$ \\
\hline
\end{tabular}
\qquad 
\begin{tabular}{|c|c|}
\hline
FEMALE GUILTY CONVICTED & $375$ \\
\hline
FEMALE NOT GUILTY CONVICTED & $350$ \\
\hline
\end{tabular}
}
\end{center}

\smallskip

We notice two things. First of all, given a single morally arbitrary distinction (male and female), it is possible to achieve group-fairness without undermining justice \emph{completely}. That is, innocent individuals (in either group) are still more likely to be acquitted than guilty ones. Second, we notice that, among women who are convicted, the share of guilty ones is little more than one half, indicating a great degree of injustice. This injustice is also significantly greater in the case of women, than in the case of men, where out of $1900$ convictions, only $400$ are mistaken. This is remarkable given that we have characterized group-fairness and justice on the basis of the same theory of $J$. Namely, all and only innocent defendants should ideally be acquitted. The example shows that, even starting from a single, monist, conception of justice, complex trade-offs may emerge between the degree to which justice is achieved and the group fairness of the imperfect procedures designed to achieve it.
\end{example}

\section{A probabilistic representation of group fairness and justice}
\label{Group fairness}
\subsection{The ROC space}

To provide an intuitive and rigorous characterization of some moral features of procedures, let us represent all the logically possible procedures that could be used to allocate utility $U$ to individuals with attributes of merit $J$. Since we are interested in the distribution of $U$ and merit $J$ (as both are essential to the above characterization of justice and fairness), we shall use a comprehensive graphical representation of all procedures in terms of how they affect the probability that $U$ gets assigned to individuals characterized (at this stage) uniquely by their attribute $J$, namely an adjustment of the well known ROC space.\footnote{The \emph{receiver operating characteristics} (ROC) graph is an illustrative method that provide a pictorial diagnostic reliability of a binary classifier system. For a more detailed analysis on such as a technique, we suggest to see \cite{Fawcett2006}.}

\begin{figure}[ht!]
\centering
\begin{tikzpicture}[rotate=-45,scale=1,>=latex]
  \fill (0,0) circle[radius=2pt];
  \fill (3,3) circle[radius=2pt];
  \fill (-3,3) circle[radius=2pt];
  \fill (0,6) circle[radius=2pt];
  \draw[thick] (0,0) -- (3,3);
  \draw[thick] (0,0) -- (-3,3);
  \draw[thick] (-3,3) -- (0,6);
  \draw[thick] (3,3) -- (0,6);
  \draw[thick, densely dotted] (0,0) -- (0,6);
    \node[scale=1,rotate=90] at (-1.7,1.3) {$\mathbb{P}(U=0 \ | \ J=1)$};
    \node[scale=1] at (1.7,1.3) {$\mathbb{P}(U=0 \ | \ J=0)$};
    \node[scale=0.75] at (2,1) {(True positive rate)};
    \node[scale=0.75, rotate=90] at (-2,1) {(False positive rate)};
    \node[scale=1.5] at (-1,3) {$S_2$};
    \node[scale=1.5] at (1,3) {$S_1$};
	\node[left, scale=1.5] at (-0.1,0) {$O$};
	\node[left, scale=1] at (-0.1,5) {$a$};
	\node[right, scale=1.5] at (-0.1,6) {$Q$};
	\node[left, scale=1.5] at (-3,3) {$B$};
	\node[right, scale=1.5] at (3,3) {$A$};
\end{tikzpicture}
\caption{The ROC space. There are several interesting objects in the picture, which deserve particular attention. We describe them in relation to our example of a procedure for convicting or acquitting a person charged with a crime in criminal law. The segment $a$ represents the line of a fully merit-insensitive procedure. $O$ is the free-for-all point, in which both the probabilities defining the procedure $\mathbb{P}(U= 0 \ | \ J=1)$ and $\mathbb{P}(U= 0 \ | \ J=0)$ are equal to $0$. $Q$, on the contrary is all-in-jail point, in which both probabilities defining the procedure $\mathbb{P}(U= 0 \ | \ J=1)$ and $\mathbb{P}(U= 0 \ | \ J=0)$ are equal to $1$. With respect to the segment $a$, the diagram is actually symmetric. The triangle  $S_1$  gives the reasonable procedures (in terms of guilty/justice), and  $S_2$  characterizes fully unreasonable procedures).}
\label{fig:procedurediamond}
\end{figure}
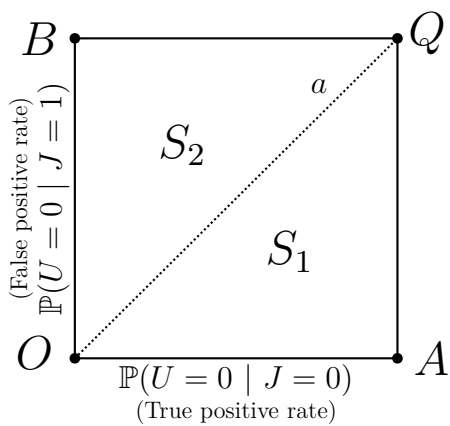
Figure \ref{fig:procedurediamond} represents - in a very intuitive way - the whole spectrum of possibilities that involves a procedure, as defined above. Any point in the diagram provides a complete description of the procedure by assigning a value (between $0$ and $1$) to $\mathbb{P}(U= 0 \ | \ J=1)$ and $\mathbb{P}(U= 0 \ | \ J=0)$. That value represents the probability for an individual to receive a certain treatment given the merit feature that individual has.  

\begin{remark}
We notice that the probability $\mathbb{P}(U= 0 \ | \ J=1)$, which defines the probability to be declared guilty and convicted if innocent, actually represents the false positive rate, and the probability $\mathbb{P}(U= 0 \ | \ J=0)$, which defines the probability to be declared guilty and convicted if (really) guilty, actually represents the true positive rate. Sometimes the graph is represented by reversing the $x$ axis with the $y$ axis. 
\end{remark}

\begin{remark}
We highlight that the use of the ROC space is quite general. Given \eqref{XU} utility $U$ could be replaced by the criterion $X$. That means that the diagram can also be used to represent the probability for an individual to fulfill the criterion deciding the treatment, given that the treatment is deserved (or not). Moreover, as the merit $J$ could be replaced by some arbitrary group $G$, the ROC space can be used to represent the probability of a given treatment, given the group.
\end{remark}

\subsection{The perfectly just procedure}

Intuitively, a procedure is perfect if, every time it is implemented, the distribution of the outcomes $U$ is perfectly justified by the theory of justice that is relevant to that domain. As proposed by \cite{loi_philosophical_2019}, a perfect procedure can be characterized as one that distributes $U$ equally between individuals who deserve the same treatment, and unequally between individuals who deserve a different treatment. It follows as a matter of definition that a perfect procedure is one that contains no inequality in $U$ among people equal in their merit, namely $J$.  

In probabilistic terms given a binary merit attribute ($J\in\{0,1\}$) and a binary disadvantage attribute ($U\in\{0,1\}$) a perfect procedure is a procedure that guarantees that 
\begin{align*}
    \mathbb{P}(U&= 0 \ | \ J=0)=1;\\
\mathbb{P}(U&= 0 \ | \  J=1)=0.
\end{align*}
    
For example, those who deserve a punishment always get punished; those who deserve no punishment are never punished.
It’s very important to highlight that for a binary utility $U\in\{0,1\}$ the diagrams (constructed for $U=0$, i.e., the distribution of convictions) already represent the same procedures described in relation to $U=1$ (i.e., to the distribution of acquittals). In fact, the following relations hold:
\begin{align*}
\mathbb{P}(U= 0 \ | \ J=0)&=1-\mathbb{P}(U=1 \ | \ J=0);\\
\mathbb{P}(U= 0 \ | \ J=1)&=1-\mathbb{P}(U=1 \ | \ J=1).
\end{align*}
In our ROC space, this is point $A$. The symmetric point $B$ that we omit from our argument gives the opposite, "perfectly unjust", procedure, in which all and only the innocent people are convicted.
\subsection{Group-fair procedures}
\label{GroupFair}
We define a procedure to be group-fair, and for the sake of brevity, "fair", in what follows, when individuals who deserve the same treatment are not  more likely to receive a favorable treatment because of the group to which they belong. This is equivalent to requiring that morally arbitrary (i.e., non $J$) features be statistically irrelevant to the distribution of the benefit or harm assigned by the procedure, $U$. For example, if being a woman is a morally arbitrary feature relative to criminal punishment, a procedure is fair only if all guilty defendants have the same chances of being convicted independently of their gender, and the same holds for all the innocent ones. This formalizes mathematically the intuitive idea that a fair procedure should not favor, intentionally or unintentionally, any individual on morally arbitrary grounds, i.e., its membership to a morally arbitrary group. Notice that fairness does not require that the probability of conviction/acquittal be the same for individuals who differ in $J$, i.e., it is not unfair for innocents and guilty defendants to have \emph{different} chances of conviction or acquittal. 
 
Mathematically speaking, fairness relative to two groups $G_1$ and $G_2$ obtains if and only if:
\begin{align*}
    \mathbb{P}(U= 0 \ | \ J=0 \wedge G_1) &= \mathbb{P}(U= 0 \ | \ J=0 \wedge G_2);\\
     \mathbb{P}(U= 0 \ | \ J=1 \wedge G_1) &= \mathbb{P}(U= 0 \ | \ J=0 \wedge G_1).
\end{align*}
A procedure is \emph{absolutely} group-fair, that is to say, fair with respect all arbitrary groups, if and only if the probability of receiving disadvantage, $U$, for those who are equally deserving, $J$, is statistically independent of membership to any possible morally arbitrary group, $G_n$ (with $n$ integer number).

\begin{remark}
We underline here that a group could be defined also by a singular individual, which constitutes a singleton (namely a set with exactly one element). 
\end{remark}

\subsection{Kinds of non-perfectly-just procedures}
\label{4.4}
A procedure can fail to be perfectly just in two main different ways. 

First, procedures can be unreasonably unjust. We define a procedure to be unreasonably unjust if it is inclined to deliver the opposite of what the individual deserves. Unjust procedures occupy the left portion of the ROC space. They are more likely to convict an innocent than a guilty defendant and more likely to acquit a guilty than an innocent defendant. 

The perfectly unreasonably unjust procedure is, as already anticipated, the point $B$, symmetric to $A$ in the ROC space. This is the point in which all and only the innocent people are convicted. If one were able to implement a perfectly just procedure, the perfectly unreasonably unjust procedure could be achieved by inverting the values of $U$ it attributes.

Second, some procedures are not unreasonably unjust, but \emph{imperfectly just}. These are the procedures in the right side of the ROC space, with the exclusion of the point $A$. They are not inclined to give individuals the opposite of what they deserve, but they make errors. 

The rate of errors can vary. Imperfectly just procedures to the right of the line $QO$ in the ROC space are more likely to convict a guilty defendant than an innocent defendant and in this sense can be considered approximations of justice, rather than approximations of injustice, overall. 

There are three special kinds of imperfectly just procedures. 

\paragraph{Merit-agnostic procedures}
Some imperfectly just procedures are \emph{perfectly merit-agnostic}, that is to say, entirely insensitive to merit ($J$). In formal terms:  the distribution of outcome $U$ is independent from the feature that justifies inequality, which we call "merit" ($J$). That is 
\[
\mathbb{P}(U= 1 \ | \ J=1) = \mathbb{P}(U= 1 \ | \ J=0) = \mathbb{P}(U=1)=p_1,
\]
and\footnote{In binary cases, the second condition follows necessarily from the first.}
\[
\mathbb{P}(U= 0 \ | \ J=1) = \mathbb{P}(U= 0 \ | \ J=0)=\mathbb{P}(U=0)=p_0,
\]
where $p_0,p_1 \in (0,1)$.

In the example of the criminal procedure, a procedure that is perfectly merit-agnostic is equally likely to convict or acquit an individual who is innocent or guilty. This can be achieved, for example, by tossing a coin in order to decide whether the defendant should be convicted or acquitted. Coin tosses and similar "random" devices easily guarantee equal chances of acquittal for \emph{all} innocent defendants, irrespective of their arbitrary traits. But they also guarantee the same chances of acquittal between \emph{innocent and guilty} defendants, that is, they are fully agnostic with respect to $J$. In the ROC space, these procedures are represented by the line $QO$.

Merit-agnostic procedures are not unreasonably unjust and they can also be group fair, since individual outcome prospects are identical across groups, for people with the same merit features. Consider assigning convictions by tossing an unbiased coin. This procedure ignores merit entirely, which in this case is the defendant's innocence or guilt. But it is also group-fair, as innocent individuals in every possible morally arbitrary group have exactly 50percent chances on average to be acquitted. Random procedures are the prime example of procedures that \emph{could} be merit agnostic, under a given description.\footnote{We use "random" here to mean a procedure that is not "deterministic" in the sense explained in Section \ref{Concepts}.} One may argue that  deterministic procedures could be merit-agnostic in special cases. We will not discuss this case in what follows because it is irrelevant to our argument. For, if a merit-agnostic procedure has to be fair to all groups, including singletons, it cannot be deterministic. (As we will argue next.)

\paragraph{Degenerate procedures}A special case of a perfectly merit-agnostic procedure is the procedure that assigns the same outcome to every individual. The all-in-jail and free-for-all points, $Q$ and $O$, in the ROC space satisfy this definition. These are procedures only in a degenerate sense, since they do not contemplate any option. 

\paragraph{Semi-perfect procedures}
Finally, there could be procedures that are perfectly just for some values of $J$, but not for others. In our binary example, this would be a procedure guaranteeing that no innocent person be convicted, but that wrongly acquits some guilty defendants. This would be perfect for innocents, but not for the guilty. Conversely, a procedure could be perfect for the guilty but not for innocents. That is, it could guarantee that all guilty defendants are convicted, but also convict some innocents.

\section{Thesis and short argument}
\label{Argument1}

Our thesis is that if a procedure is deterministic, it is either perfect or there is at least one group, that is morally arbitrary, such that probabilities of obtaining the deserved treatment differ on average between the in- and the out- groups (i.e., group fairness is violated).

The argument can be summarized as follows. Suppose we are dealing with a deterministic procedure, where we can always conceive the set of determinant facts, $X$, causing the acquittal of the defendant. These facts will either obtain, or not. If, when the facts obtain, the defendant is innocent, the procedure is \emph{perfect}. Otherwise, the determinant facts will not obtain for all innocent individual, or they will also obtain for some guilty ones. In this case, the procedure is imperfect, and the two groups $X=0$ and $X=1$ are morally arbitrary. Innocent individuals have different chances of benefiting from the procedure, depending on their two groups $X=0$ or $X=1$. Therefore, group fairness is not satisfied. This argument does not consider random procedures, those for which a set of determinant facts, $X$, cannot be defined. We therefore conclude that, of imperfect procedures, \emph{only} random procedures, can be group fair \emph{absolutely}. If acquittal is decided by the throw of a coin, for instance, there is no set of facts (about the defendant and the circumstances of the case) prior to the throw itself which determines whether the defendant will be acquitted or convicted. No $X$ can be defined, which blocks our argument.\footnote{If a procedure includes a randomizing mechanism, such heads or tail, and all the causes that cause a coin to land head or tail, in a deterministic world, are included in the description of a procedure initial state, so that a certain outcome follows necessarily in standard operating conditions, then that procedure counts as deterministic too.}

It may be objected that membership to $X=1$ is not morally arbitrary. Even if $X=1$ does not track innocence perfectly, it is also not unrelated to it entirely. If $X=1$ is a reasonably good criterion for acquittals, as it must be if we are in (vii), then $X=1$ tracks innocence to a certain degree. That is, defendants for whom $X=1$ are more likely to be innocent than defendants for whom $X=0$.
The rejoinder is to observe that, if someone were to use sex as the determinant criterion for a decision, \emph{even in a world in which it were reasonably well correlated with the just outcome} this would still strike most people as unfair. The philosophical challenge is therefore to differentiate the obviously unfair criterion of sex (even when it can be correlated with the just outcome) from any other type of $X$. 

In Section \ref{Discussion} we will explain why this result is important from a moral and political point of view and should affect all subsequent theorizing about fairness in relation to imperfect procedures.

\section{Detailed argument}
\label{Argument2}
We shall now provide the detailed argument.
Here it is useful to keep in mind the ROC space that makes our line of reasoning easier. Let $X$ be a criterion and let us suppose the following logic equivalence between utility $U$ and criterion $X$ such that $X$ is a necessary and sufficient condition for $U$. Thus, we immediately have 
\begin{align}
X= 1 \iff& U= 1; \label{XsseU_1}\\
\label{XsseU_0}
X=0 \iff& U=0.
\end{align}

Now let us consider the following generic condition for a (binary) procedure:
\begin{align*}
\mathbb{P}(U=0  \ | \ J=1)&=k;\\
\mathbb{P}(U= 0  \ | \ J=0)&=h;\\
\mathbb{P}(U= 1 \ | \ J=1)&=1-k=p;\\
\mathbb{P}(U= 1 \ | \ J=0)&=1-h=q.
\end{align*}
Here $k$ and $h$ are the probabilities that someone be acquitted if one is, respectively, innocent or guilty of the crime charged to her. The conditions $p=1-k$ and $q=1-h$ can be deduced directly from our binary hypothesis (i.e., one can deduce the third and the fourth equation directly from the first and the second).
We can divide the triangle $S_1$ of Figure \ref{fig:procedurediamond} in its fundamental components distinguishing seven cases (three vertices, three edges, and the interior surface). These seven possibilities represent all the logical possibilities (we highlight, once again, that the triangle $S_2$ of Figure \ref{fig:procedurediamond} represents the perversion of justice, with all its possible degrees of imperfection, and can be easily recovered by symmetry, but we are not considering it in our argument). We remind the reader that $h$ is the probability of conviction for the guilty defendant and $h$ for the innocent.

\paragraph{Vertices}

\begin{itemize}
    \item[(i)]  \emph{Perfectly just procedure}: $h=1$ and $k=0$.
    \item[(ii)] \emph{Everyone convicted}: $h=1$ and $k=1$.
    \item[(iii)] \emph{Everyone acquitted}: $h=0$ and  $k=0$.
\end{itemize}
\paragraph{Edges}

\begin{itemize}
    \item[(iv)] \emph{Perfect for guilty defendants}: $h=1$ and $k \in (0,1)$.
    \item[(v)] \emph{Perfect for innocent defendants}: $k=0$ and $h \in (0,1)$.
    \item[(vi)] \emph{Perfectly merit-agnostic}: $k=h$ with $k\in (0,1)$ and $h\in(0,1)$. We are in the segment $a$ (the dots line) of the diagram (extremes excluded).
\end{itemize}
\paragraph{Interior surface}

\begin{itemize}
    \item[(vii)]  \emph{Generic imperfect procedures}: $h>k$ (and $h\neq0$, $k\neq1$ ). We are in the interior of the triangle $S_1$ (i.e., edges excluded).
\end{itemize}

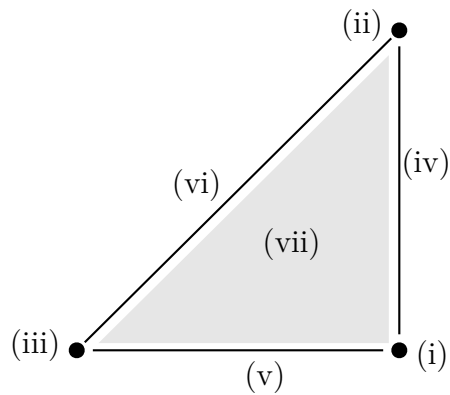
\begin{figure}[ht!]
\centering
\begin{tikzpicture}[rotate=-45,scale=1,>=latex]
  \fill[scale=0.9,fill=gray!20,shift={(0.15,0.3)}] (0,0)--(0,6)--(3,3);
  \fill (0,0) circle[radius=3pt];
  \fill (3,3) circle[radius=3pt];
  \fill (0,6) circle[radius=3pt];
  \draw[thick] (0.15,0.15) -- (2.85,2.85);
  \draw[thick] (3-0.15,3+0.15) -- (0.15,5.85);
  \draw[thick] (0,0.15) -- (0,5.85);
    \node[scale=1] at (2,1.5) {(v)};
     \node[scale=1] at (1.5,5) {(iv)};
    \node[scale=1] at (1,3) {(vii)};
	\node[left, scale=1] at (-0.1,0) {(iii)};
	\node[right, scale=1] at (3.1,3) {(i)};
	\node[left, scale=1] at (-0.1,3) {(vi)};
	\node[left, scale=1] at (-0.1,6) {(ii)};
\end{tikzpicture}
\end{figure}

 \paragraph{Perfect procedures}Let us first of all suppose that a procedure is perfect. Then, by definition all people obtain what they deserve. That means that all and only the guilty people ($J=0$) are convicted ($U=0$). Since the procedure reaches the outcome $U=0$ (conviction) when and only when there is sufficient evidence of the crime ($X=0$), a perfect procedure is also one in which the criterion for acquittal is never satisfied in the case of every actual crime ($X=0 \iff J=0$) and it is always satisfied when the defendant was actually not responsible for it ($X=1 \iff J=1$). Or in other words, the criterion for acquittal perfectly tracks the facts of the crimes and the merit of the defendant.

This procedure is perfectly fair because it ensures that people with the same value of $J$ have the same outcome, hence their probabilities to obtain those outcomes are the same. Perfectly just procedure are fair, but unfeasible.

\paragraph{Merit-agnostic procedures} 
By definition of (vi), (ii) and (ii) the procedure is insensitive to differences in merit $J$. As explained above, it is important to highlight that group fairness obtains \emph{absolutely} only if it obtains for all logically possible morally arbitrary groups. This includes those groups containing a single individual. It is trivial to notice that the person belonging to the singleton can be either innocent ($J=1$) or guilty ($J=0$), at least in our binary model. Indeed it follows immediately by definition that these procedures can be described with $\mathbb{P}(U=0)=p_0$, and $\mathbb{P}(U=1)=p_1$. This induces the independence to any \emph{a priori} criterion.\footnote{We exclude the \textit{a posteriori} criterion of the lottery procedure $X=U$, e.g., the winners of a lottery, where $X$ is not the initial state of the procedure, but coincides with its outcome.} For simplicity, suppose that the population consists of several groups, each of which contains one and only one individual (e.g., there is only one person at a given intersection of gender, race, genetic, and all other attributes), each of which is either innocent or guilty. A \emph{deterministic} procedure can only be merit-agnostic if it acquits the single individual in a certain proportion (say, $1/3$) of groups of innocent defendants and if it acquits the individual in the same proportion of groups of guilty ones. Clearly, this violates group fairness because there are morally arbitrary groups in which innocents are certainly acquitted ($1/3$) and others in which they are certainly convicted  ($2/3$). The example generalizes.
\smallskip

Let us consider all other cases: namely (v), (vi) and (vii). 
\paragraph{Imperfectly just procedures} Let us begin with (vii). Mathematically speaking, the argument for (vii) can be formalized, without loss of generality, as follows. Let us consider $h>k$, namely
\begin{equation*}
\mathbb{P}(U= 0  \ | \ J=0)>\mathbb{P}(U=0  \ | \ J=1).
\end{equation*}
However, by using condition \eqref{XsseU_0}, we have immediately
\begin{equation}
\label{X01}
\mathbb{P}(U=0  \ | \ J=0 \wedge X=0)=1=\mathbb{P}(U=0  \ | \ J=1 \wedge X=0),
\end{equation}
and
\begin{equation}
\label{X10}
\mathbb{P}(U=0  \ | \ J=0 \wedge X=1)=0=\mathbb{P}(U=0  \ | \ J=1 \wedge X=1).
\end{equation}
By these two equations the role of $X$ is fulfilled: the original procedure \emph{collapses} in the two possible results provided by \eqref{X01} and \eqref{X10}. Indeed, the criterion $X$, by definition, represents a necessary and sufficient condition for the outcome $U$. This implies directly that acquittal and conviction are entirely independent of $J$ and fully determined by $X$. It follows that individuals with the same values of $J$ have different chances depending on the value of $X$. Thus, the groups identified by the criterion $X$ (we recall that $X=0$ and $X=1$ are complementary in the binary case) actually identify an unfair procedure. 
\paragraph{Semi-perfect procedures} Let us now consider the special cases (v) and (vi). The procedures in (v) are fair to the innocents: since all innocents are certain to be acquitted, it trivially follows that innocent defendants of different groups have the same chances. But these procedures are not fair to the guilty, since guilty individuals have different chances of conviction depending on their membership to $X=1$. \emph{Mutatis mutandis}, it should be clear that the procedures in (iv) are fair to the guilty, but not to the innocents.\footnote{This is not to say that the procedure is insensitive to merit, for, by assumption, the imperfect procedures in question are such that they gives guilty defendants higher chances of conviction than innocent ones, that is $h>k$.} We underline that a similar argument can be provided for $U=1$ by using condition \eqref{XsseU_1}.

\section{Discussion}
\label{Discussion}

Ideally, group-fairness should constrain the way an imperfect, deterministic procedure achieves (a given degree) of justice. This seems possible when the number of group regarded as morally arbitrary is limited, as in the Example \ref{exampleMF}. However, if one requires the fulfillment of group fairness with respect to \emph{all} morally arbitrary groups, this can only be achieved by avoiding determinism. In a non-deterministic procedure, a given initial state of the procedure may correspond to some outcomes in one case, and to other outcomes in another, without any change in its inputs,\footnote{Or conditions, or circumstances, that can be modelled as the initial state of the procedure.} in normal operating conditions. This is problematic since this random element can be seen as the opposite of what justice should try to achieve where there is a clear moral ground (e.g., innocence) to justify an outcome.

One could take this as a reason to reject group-fairness altogether. If so, when assessing procedures that are imperfect, one simply ought to minimize injustice, disregarding group-distributive effects altogether. A less radical option is to amend our account of group-fairness. This requires a different theory of what makes \emph{certain} groups "morally arbitrary". "Different" here means that the concept of the morally arbitrary is not reducible to "something other than what, morally speaking, justifies inequality". This is the direction in which we want to push our argument, philosophically. We believe our result is best interpreted as an argument supporting the quest for such philosophical view.

We are aware that we make certain assumptions about the nature of procedures that are quite restrictive. We hope this is seen as a useful attempt to describe the nature of at least an important kind of procedure. 
\section{Conclusions}
In this paper we have provided a mathematical proof that a decidable procedure cannot fulfill group-fairness in relation to all possible relevant groups. Arguably this result generalizes to all procedures that are deterministic under certain descriptions.

Group justice, as defined, bears significant analogies to certain measures of fairness in the machine learning literature, in particular equality in the false positive/negative rates of different groups. If our argument is correct, group fairness can be defined in a similar way in relation to justice. This shows that the relation between group fairness, understood as quality of imperfect procedures, and justice, is normally problematic.

Hopefully our paper will motivate other scholars to study questions of group-fairness in imperfect procedures, regarded as a general type of procedures, conceptually broader than statistical models.
\label{Conclusions}

\bibliographystyle{plain}  
\bibliography{bibliography}

\begin{thebibliography}{10}

\bibitem{arneson_equality_1989}
Richard~J. Arneson.
\newblock Equality and {Equality} of {Opportunity} for {Welfare}.
\newblock {\em Philosophical Studies}, 56(1):77--93, 1989.

\bibitem{barocas_fairness_nodate}
Solon Barocas, Moritz Hardt, and Arvind Narayanan.
\newblock \emph{Fairness in {Machine} {Learning}: {Limitations} and
  {Opportunities}}.
\newblock \url{https://fairmlbook.org}.

\bibitem{brouwer_why_2018}
Huub Brouwer and Thomas Mulligan.
\newblock Why not be a desertist?
\newblock {\em Philosophical Studies}, June 2018.

\bibitem{ceva_interactive_2016}
Emanuela Ceva.
\newblock {\em Interactive {Justice}: {A} {Proceduralist} {Approach} to {Value}
  {Conflict} in {Politics}}.
\newblock Routledge, May 2016.

\bibitem{cohen_currency_1989}
Gerald~Allan Cohen.
\newblock On the {Currency} of {Egalitarian} {Justice}.
\newblock {\em Ethics}, 99(4):906--944, 1989.

\bibitem{daniels_accountability_2000}
Norman Daniels.
\newblock Accountability for reasonableness.
\newblock {\em BMJ : British Medical Journal}, 321(7272):1300--1301, November
  2000.

\bibitem{di_bello_profile_2020}
Marcello Di~Bello and Collin O’Neil.
\newblock Profile {Evidence}, {Fairness}, and the {Risks} of {Mistaken}
  {Convictions}.
\newblock {\em Ethics}, 130(2):147--178, January 2020.
\newblock Publisher: The University of Chicago Press.

\bibitem{Fawcett2006}
Tom Fawcett.
\newblock An introduction to roc analysis.
\newblock {\em Pattern Recognition Letters}, 27(8):861--874, 2006.

\bibitem{hardt_equality_2016}
Moritz Hardt, Eric Price, and Nati Srebro.
\newblock Equality of opportunity in supervised learning.
\newblock In {\em Advances in neural information processing systems}, pages
  3315--3323, 2016.

\bibitem{heidari_moral_2019}
Hoda Heidari, Michele Loi, Krishna~P. Gummadi, and Andreas Krause.
\newblock A {Moral} {Framework} for {Understanding} {Fair} {ML} {Through}
  {Economic} {Models} of {Equality} of {Opportunity}.
\newblock In {\em Proceedings of the {Conference} on {Fairness},
  {Accountability}, and {Transparency}}, {FAT}* '19, pages 181--190, New York,
  NY, USA, 2019. ACM.
\newblock event-place: Atlanta, GA, USA.

\bibitem{herlitz_measuring_2016}
Anders Herlitz and David Horan.
\newblock Measuring needs for priority setting in healthcare planning and
  policy.
\newblock {\em Social Science and Medicine}, 157:96--102, 2016.

\bibitem{loi_philosophical_2019}
Michele Loi, Anders Herlitz, and Hoda Heidari.
\newblock A philosophical theory of fairness for prediction-based decisions.
\newblock {SSRN} {Scholarly} {Paper} ID 3450300, Social Science Research
  Network, Rochester, NY, September 2019.

\bibitem{loi_fair_2021}
Michele Loi, Anders Herlitz, and Hoda Heidari.
\newblock Fair {Equality} of {Chances} for {Prediction}-based {Decisions}.
\newblock In {\em Proceedings of the 2021 {AAAI}/{ACM} {Conference} on {AI},
  {Ethics}, and {Society}}, {AIES} '21, page 756, New York, NY, USA, July 2021.
  Association for Computing Machinery.

\bibitem{miller_principles_1999}
David Miller.
\newblock {\em Principles of {Social} {Justice}}.
\newblock Harvard University Press, Cambridge, Mass, 1999.

\bibitem{rawls_theory_1971}
John Rawls.
\newblock {\em A {Theory} of {Justice}}.
\newblock Harvard University Press, Cambridge, MA, 1 edition, 1971.

\bibitem{sen_equality_1982}
Amartya Sen.
\newblock Equality of {What}?
\newblock In {\em Choice, {Wefare} and {Measurement}}, pages 353--369.
  Cambridge University Press, Cambridge MA, 1982.

\bibitem{temkin_equality_2017}
Larry Temkin.
\newblock Equality as {Comparative} {Fairness}.
\newblock {\em Journal of Applied Philosophy}, 34(1):43--60, February 2017.
\newblock WOS:000397192700006.

\bibitem{walzer_spheres_1983}
Michael Walzer.
\newblock {\em Spheres of {Justice}: {A} {Defense} of {Pluralism} and
  {Equality}}.
\newblock Basic Books, New York, 1983.

\bibitem{wong_democratizing_2019}
Pak-Hang Wong.
\newblock Democratizing {Algorithmic} {Fairness}.
\newblock {\em Philosophy \& Technology}, June 2019.

\end{thebibliography}

\end{document}